
\magnification=1200
\def\em{{\bf M}}
\def\trans{^{\rm T}}
\def\Mab{M_{\bar a}^b}
\def\gac{g_{\bar a c}}
\def\gab{g_{\bar a b}}
\def\di{\partial_i}
\def\sinh{\rm sinh}
\def\cosh{\rm cosh}
\hfill INFN-8/92-DSF

\hfill hep-th/9206037
\vskip 3cm
\centerline{ \bf GEOMETRY AND INTEGRABILITY}
\vskip .5cm
\centerline{\bf OF TOPOLOGICAL-ANTITOPOLOGICAL
FUSION }

\vskip 1cm

\centerline{B.DUBROVIN\footnote{$^\dagger$}{On leave of abscence from
Dept. Mech. \& Math.,
Moscow State University,
119899, Moscow}}
\vskip 1cm
\centerline {I.N.F.N., Sez. di Napoli}
\centerline{Mostra d'Oltremare, Pad.19}
\centerline{80125 NAPOLI, Italy}
\centerline{E-mail DUBROVIN@NA.INFN.IT}
\vskip 1cm

{\bf Abstract.}

Integrability of equations of topological-antitopological fusion
(being proposed by Cecotti and Vafa) describing ground state metric
on given 2D topological field theory (TFT) model, is proved. For massive TFT
models these equations are reduced to a universal form
(being independent on the given TFT model) by gauge transformations.
For massive perturbations of topological conformal field theory
models the separatrix solutions of the equations bounded at infinity
are found by the isomonodromy
deformations method. Also it is shown that ground state metric together
with some part of the underlined
 TFT structure can be parametrized by pluriharmonic maps of the
coupling space to the symmetric space of real positive definite
quadratic forms.

\vfill\eject

{\bf Introduction.}
\medskip

The idea of topological field theories (TFT) as solvable models without local,
propagating degrees of freedom was proposed in [1].
In [1, 2,
3, 4] it was shown that topological correlators (at tree level) in a 2D
TFT model
are holomorphic functions on moduli of the TFT
model obeying an overdetermined system
of nonlinear PDE (the equations of associativity of primary operator
algebra). Integrability of this equations was proved in [5].

The problem of calculation of the ground state metric of a family of
TFT was studied in general situation (for both massless and massive
theories) in [6].
In this paper a system of PDE
for the ground state metric (being a Hermitian metric on the moduli space
of TFT) was derived. The topological and "antitopological" (i.e.
complex conjugate) correlators serve as coefficients of these PDE.
This general construction of calculating of ground state metric was
called in [6] a {\it topological-antitopological fusion}.
The equation of the same form arise for the metric on moduli
space of Calabi-Yau varieties [7, 8].
The Hermitian metric on the moduli
space in this case is the same as the Zamolodchikov metric [9]
of the underlying $N=2$ superconformal field theories. In this case the
metric is K\"ahler having special properties. The geometry of the moduli
space with this metric is called in [7, 8] {\it special geometry}.
Thus general solutions of the equations of [6] can be called also
{\it generalized special geometry}. Also in [6, 10, 11]
a number of
particular integrable reductions of the main equations was found. It was
shown that under some symmetry assumption
the equations of topological-antitopological fusion can be reduced to affine
Toda equations (particularly, to Euclidean sinh-Gordon) and to some
other integrable systems of the soliton theory. For massive
perturbations of topological conformal field theory (TCFT) many
particular reductions of the main equations can be solved via
the Painlev\'e transcendents of the third kind. More complicated
reduction of the  equations of topological-antitopological fusion
was investigated numerically in [12].

The present paper can be considered as continuation of the
investigations having been started in [5]
of the r\^ole of integrable systems in the TFT. The main
aim of this paper is to prove integrability of the equations
of topological-antitopological fusion in the general case.
This integrability immediately follows from the zero-curvature
representation of these equations depending on a spectral parameter
being obtained in Sect.1.

In Sect.2 it was proved for massive TFT models that
the equations of topological-antitopological fusion  can be reduced to a
universal integrable PDE system with constant coefficients (i.e. not
depending on the given TFT model). For models with two
primaries this system coincides with the Euclidean sinh-Gordon.
For massive perturbations of TCFT the ground state metric can be found from
the equations (generalizing the Painlev\'e III) of isomonodromy deformations
of a linear operator with rational coefficients
(again this operator is universal, i.e. it does not depend
on the concrete TCFT model). The separatrix solutions of these
equations satisfying at the infinity the boundary conditions of [6] are
found in Sect.2 using Riemann boundary value problem machinery.

A nice geometrical reformulation of
the equations of topological-antitopological fusion is given in Sect.3.
It is shown that any solution of these equations determines a
pluriharmonic map (i.e. harmonic along complex directions) of the moduli
space of the TFT model to the symmetric space of real positive definite
quadratic forms (in fact a loop in the space of pluriharmonic maps).
Conversely, any such a map determines a family of
topological-antitopological fusion structures on the moduli space together
with the underlying TFT structure. Functional parameters of the
family can be described explicitly
in differential geometric terms. This relation to the theory of harmonic
maps probably
can be useful to describe possible topo\-logical-anti\-topo\-logical fusion
structures "in large" (i.e. using an appropriate information about
topology of the moduli space of the TFT model).
\medskip
{\bf 1.Zero-curvature representation for the equations
of topological-anti\-topo\-lo\-gical fusion.}
\medskip

Let $\em$ be a complex manifold of (complex) dimension $n$ with
a nondegenerate
holomorphic complex quadratic form
$$\eta = \eta_{ab}(z)dz^adz^b,~ \det(\eta_{ab})\neq0 \eqno(1.1)$$
and a Hermitian positive definite form
$$g=g_{\bar a b}(z,\bar z)d\bar z^adz^b,~ \overline{g_{\bar b a}}=
g_{\bar a b},
{}~\det (g_{\bar a b})\neq 0 \eqno(1.2)$$
(the bar means complex conjugations).

{\bf Definition 1.1.} The pair $\eta, g$ is called {\it compatible} if
there exists a complex connection $D = (\Gamma_{ab}^c, \Gamma_{\bar a \bar b}
^{\bar c} =\overline{ \Gamma_{ab}^c})$, where for any complex vector field
$X=X^a\partial_a$
$$D_cX^a=\partial_cX^a+\Gamma_{cb}^aX^b, \eqno(1.3a)$$
$$D_{\bar c}X^a=\bar{\partial}_cX^a, \eqno(1.3b)$$
$$D_{\bar c}=\bar{D_c} \eqno(1.3c)$$
such that
$$D_c\eta_{ab}\equiv \partial_c\eta_{ab}-
\Gamma_{ca}^d\eta_{db}-\Gamma_{cb}^d\eta_{db}=0 \eqno(1.4a)$$
$$D_cg_{\bar a b}\equiv
\partial_cg_{\bar a b}-\Gamma_{cb}^dg_{\bar a d}=0. \eqno(1.4b)$$

Note that the equations $D_{\bar c}\overline{\eta_{ab}}=0$,
$ D_{\bar c}g_{\bar a b}=0$
follow from (1.4).

It is clear that the connection $D$ for a compatible pair $\eta, g$ is
determined uniquely. The matrices $\Gamma_c=(\Gamma_{ca}^b)$ have the form
$$\Gamma_c=g^{-1}\partial_c g,~ g=(g_{\bar a b}). \eqno(1.5)$$

It immediately follows from the definition that the tensor $M=(M_{\bar a}^b)$

$$\Mab=\gac \eta^{cb},~ (\eta^{cb})=(\eta_{cb})^{-1} \eqno(1.6)$$
obeys the equations $$M{\bar M}= const .1. \eqno(1.7)$$
The compatible pair $\eta,~ g$ is called {\it normalized} if
$$M{\bar M} = 1. \eqno(1.8)$$

On a complex manifold with a normalized compatible pair $\eta, ~g$
there is a canonical ${\bf C}$-linear isomorphism between the spaces of
complex tensors of the type $\left(_{q~q'}^{p~p'}\right )$ (i.e., tensors
with the components of the form
$T_{j_1...j_q\bar l_1...\bar l_{q'}}^{i_1...i_p \bar k_1...\bar k_{p'}}$ -
the notations as in [13]) for given $p+p'+q+q'$.
In other words, the operations of raising and lowering of indices via
$\eta_{ab}
,~\overline{ \eta_{ab}},~$and
 $\gab$ commute. The parallel transport being specified by the connection
$\Gamma$ respects this isomorphism. Also an anticomplex involution $\tau$
acting on the (complexified) tangent space
$T\em \otimes {\bf C} = T^{1,0} \em \oplus T^{0,1}\em$ is defined as follows:
$$\tau (X^a\partial_a+X^{\bar a}\bar{\partial}_a)=\Mab\overline
{X^a}\partial_b+\overline{\Mab}\overline{X^{\bar a}}\bar{\partial}_b,
\eqno(1.9a)$$
$$\tau(T^{1,0}\em )=T^{1,0}\em ,~ \tau(T^{0,1}\em )=T^{0,1}\em
, \eqno(1.9b)$$
$${\tau}^2=1,~ \tau(\lambda x)=\bar{\lambda}\tau (x)~ {\rm for}~ \lambda \in
\bf C. \eqno(1.9c)$$
The operator $\tau$ commutes with the standard complex conjugation
$$T^{1,0}\to T^{0,1},~ X \mapsto \bar X, \eqno(1.10a)$$
$$\tau (\bar X)= \overline{\tau (X)}. \eqno(1.10b)$$
The complex inner product
$$<X,Y> = \eta_{ab}X^a Y^b \eqno(1.11)$$
and the Hermitian scalar product
$$(X,Y) = \gab \overline{X^a} Y^b \eqno(1.12)$$
are related by the equation
$$(X,Y) = <\tau (X),Y>. \eqno(1.13)$$
The operator $\tau$ is antiorthogonal with respect to the inner product $
<~,~>$:
$$<\tau (X), \tau (Y)> = \overline{<X,Y>}. \eqno(1.14)$$
It is covariantly constant with respect to the complex connection $D$:
$$D_c \Mab \equiv \partial_c \Mab + \Gamma_{cd}^a M_{\bar b}^d = 0.
 \eqno(1.15)$$
Also one should have the condition of positive definiteness
$$<\tau (X),X> ~> ~0~~{\rm for}~X\neq 0. \eqno(1.16)$$

{\bf Proposition 1.1.} {\it All compatible normalized pairs $\eta , g$
with fixed $\eta$ are in 1-1 correspondence with anticomplex involutions
of the form (1.9 ) - (1.16 ).}

The proof is straightforward.

An anticomplex involution $\tau$ with the above properties also
will be called {\it compatible} with the complex metric $\eta$.

The group of holomorphic automorphisms $A = (A_a^b(z))$ of
$T^{1,0}\em $ acts on normalized compatible pairs as follows:
$$\eta \mapsto A^{\trans} \eta A,~ g \mapsto A^{\dagger}  g A,~
M \mapsto A^{-1} M \bar A, \eqno (1.17a)$$
$$\Gamma_a \mapsto A^{-1} \Gamma_a A + A^{-1} \partial_a A. \eqno(1.17b)$$

The conection $D$ for the compatible pair $\eta , g$ is not
 symmetric. If $\hat{\Gamma} = (\hat{\Gamma}_{ab}^c)$ is the Levi-Civita
connection for the metric $\eta$ (i.e. $\hat{\Gamma}_{ab}^c = \hat{\Gamma}
_{ba}^c,~ \hat D_c \eta_{ab}\equiv \partial_c \eta_{ab} -
\hat{\Gamma}_{ca}^d \eta_{db} - \hat{\Gamma}_{cb}^d \eta_{ad} = 0$) then
the difference
$$T_{ab}^c = \Gamma_{ab}^c - \hat{\Gamma}_{ab}^c \eqno(1.18a)$$
is a $\left(_2^1 \right)$ tensor. It obeys the symmetry
$$T_{ab}^c \eta_{cd} + T_{ad}^c \eta_{cb} = 0. \eqno(1.18b)$$
If one of the metric $\eta$ or $g$ is flat then vanishing of the tensor
 $T_{ab}^c$ (in fact, vanishing of the skew-symmetric part
$T_{[ab]}^c = \Gamma_{[ab]}^c$) is equivalent to simultaneous reducibility
via holomorphic change of coordinates of the pair $\eta , g$ to a constant
form. Note that the holomorphic part of the Riemann curvature tensor
of the connection $\Gamma_{ab}^c$ vanish:
$$\left[ D_a , D_b \right] \equiv 0. \eqno(1.19)$$

{\bf Remark.} For any anticomplex involution $\tau$ in a $n$-dimensional
complex
space $T$ there exists a $n$-dimensional $\tau$-invariant
real subspace $V\subset T$ such
that $T$ is isomorphic to the complexification of $V$
$$\tau |_V=1,~T=V\oplus iV.\eqno(1.20)$$
Indeed, let
$$V_{\pm} = \left( {1\pm \tau \over 2}\right) T.$$
One has
$$T = V_+\oplus V_-,$$
$$\tau\vert_{V_{\pm}}=\pm 1.$$
Because of the antilinearity (1.9c) we obtain
$$iV_+\subset V_-,~ iV_-\subset V_+.$$
Hence $V_-=iV_+$. Putting $V=V_+$ we obtain (1.20). If a basis of the
space $T$ is chosen in $V$ then the operator $\tau$ in this basis
is represented by the unity matrix. In other words, any matrix
$M=(\Mab )$ satisfying (1.8) can be represented in the form
$$M=\Phi \bar{\Phi}^{-1}\eqno(1.21a)$$
for some complex matrix $\Phi$. The matrices of the tensors $\eta$,
$g$ in such a basis coincide
$$G=\Phi\trans\eta\Phi = \Phi^{\dagger}g\Phi\eqno(1.21b) $$
so $G$ is a real symmetric matrix. Hence the only algebraic invariant
of a normalized pair $\eta$, $g$ is the signature of the Hermitian metric
$g$. For positive definite Hermitian metric $g$ matrices of all the
tensors $M$, $\eta$, $g$ can be reduced simultaneously to the unity matrix
(in one point of the manifold ${\bf M}$).
Globally on a quasi-Fr\"obenius manifold $\em$ the distribution of
the kernels
$$V={\rm ker}(1-\tau )\subset T^{1,0}\em\eqno(1.22a)$$
determines a real $n$-dimensional bundle $\cal V$ over $\em$ such
that
$${\cal V}\subset T^{1,0}.\eqno(1.22b)$$
One has
$$T^{1,0}={\cal V}\otimes {\bf C}.\eqno(1.22c)$$
This bundle uniquely determines the antiinvolution $\tau$ (i.e. the
tensor $\Mab$). The tensors $\eta$, $g$ specify a positive definite quadratic
form on $\cal V$. In other words, they specify a section $G$
(see (1.21b)) of the associated bundle $Q({\cal V})$ of positive definite
quadratic forms on the bundle $\cal V$.

To write the equations of topological-antitopological fusion (or,
equivalently, the generalised equations of special geometry) we need to
introduce the notion of Fr\"obenius manifold (see [5]).
I recall that a commutative associative algebra $A$ with a unity is
called {\it Fr\"obenius} if there is a nondegenerate invariant inner
product $<~,~>$ on $A$:
$$<ab,c> = <a,bc>. \eqno (1.23)$$
  $\em$ is called (complex) quasi-Fr\"obenius manifold if a structure of
Fr\"obenius algebra over the ring ${\cal F} (\em )$ of holomorphic functions on
$\em$ is fixed on the space ${\rm Vect}(\em )$
of holomorphic vector-fields. It is
assumed that the invariant inner product on ${\rm Vect}(\em )$ is specified
by a nondegenerate holomorphic quadratic form $\eta$ (see (1.1)).
In local complex coordinates the multiplication law and the inner product
should read
$$(X \cdot Y)^c(z) = X^a(z)Y^b(z) c_{ab}^c(z) \eqno(1.24a)$$
$$<X,Y> = \eta_{ab} (z)
X^a(z)Y^b(z) \eqno(1.24b)$$
where $c_{ab}^c$, $\eta_{ab}$ are holomorphic tensors on $\em$. These
satisfy the equations
$$c_{ba}^c = c_{ab}^c, \eqno(1.25a)$$
$$c_{ab}^s c_{sc}^d = c_{as}^d c_{bc}^s \eqno(1.25b)$$
$$c_{abc} \equiv c_{ab}^s \eta_{sc} = c_{acb}. \eqno(1.25c)$$
If $e=(e^a)$ is the unity (holomorphic) vector field then
$$e^a c_{ab}^c = \delta_b^c \eqno (1.26)$$
(the Kronecker delta).

A quasi-Fr\"obenius $\em$ is called {\it Fr\"obenius manifold} (see
[5]) if the curvature of the connection
$$\tilde{\nabla}_X^{(\lambda)}Y = \nabla_XY + \lambda X \cdot Y
\eqno(1.27)$$
vanish identically in the spectral parameter $\lambda$. Here
$\nabla$ is the Levi-Civita connection for $\eta$. The complex
metric $\eta$ on a Fr\"obenius manifold $\em$ is flat. That means that
in an appropriate local coordinates $t^{\alpha},~ \alpha = 1, \dots ,n$,
$\eta$ has a constant form
$$\eta = \eta_{\alpha \beta}dt^{\alpha}dt^{\beta},~ \eta_{\alpha \beta}
= {\rm const.} \eqno(1.28)$$
The structure constants $c_{\alpha \beta \gamma} (t)$ in the flat
coordinates can be represented in the form
$$c_{\alpha \beta \gamma} (t) = \partial_{\alpha} \partial_{\beta}
\partial_{\gamma}F(t) \eqno(1.29)$$
for some function $F(t)$, $\partial_{\alpha} = \partial / \partial t^{\alpha}$.
In topological field theory (TFT) with $n$ primary fields  $\phi _1, ... ,
\phi _n$ the tensors $\eta _{\alpha \beta}$ and $c_{\alpha \beta \gamma}$ are
the (tree-level) double and triple correlators of the primaries respectively.
 The coordinates $t^1, \dots ,t^n$ are the coupling constants of the perturbed
 TFT model, where the Lagrangian should be perturbed as ${\cal L} \mapsto
{\cal L} - \sum  t^{\alpha} \int \phi_{\alpha}$ (see [1 - 4] for details).
And $F(t)$ coincides with the primary free energy (at tree-level).

The associativity conditions (1.25b) read as a system of nonlinear PDE
for the function $F(t)$. This system of PDE was called in [5] as
Witten - Dijkgraaf - E.Verlinde - H.Verlinde (WDVV) system. Flatness of the
connection (1.27) gives the zero-curvature representation ("Lax pair")
depending on the spectral parameter $\lambda$ for WDVV equations.

Particularly, a (quasi-) Fr\"obenius manifold $\em$ is called
{\it massive} if the
Fr\"obenius algebra $(c_{\alpha \beta}^{\gamma}$, $ \eta_{\alpha \beta})$
is semisimple (i.e. has no nilpotents) for any $t$. Local structure
of massive
Fr\"obenius manifolds can be described using an appropriate version of
inverse spectral transform (see [5] and Sect.2  below) for WDVV.

A quasi-Fr\"obenius structure on $\em$ is called {\it integrable} if  local
coordinates $u^1, \dots , u^n$ exist such that the structure tensor
$c = (c_{ij}^k)$ in these coordinates does not depend on $u$. Particularly,
any massive Fr\"obenius manifold is integrable [5]. In other
words, {\it canonical} local coordinates $u^1, \dots , u^n$ exist on a
massive Fr\"obenius manifold $\em$ such that the law of the multiplication
(1.24a) of the corresponding basic vector fields $\partial_i = \partial
/\partial u^i$ has the form

$$\di \cdot \partial_j = \delta_{ij} \di. \eqno(1.30)$$
These coordinates are determined uniquely up to permutations and shifts.

Let us come back to arbitrary quasi-Fr\"obenius manifold
$(c_{ab}^c(z),~ \eta_{ab}(z))$. I am going to define (following
[6]) special geometry structure on a given quasi-Fr\"obenius
manifold. If $\em$ is a Fr\"obenius manifold (i.e. a TFT model) then
these special geometry structures on $\em$ are called also as
topological-antitopological fusions of the given TFT model [6].

Let us denote by $C_{a}$ the operators
$$C_{a}=(c_{ab}^c(z)). \eqno(1.31)$$

{\bf Definition 1.2.} (see [6]). A compatible pair $\eta ,~ g$
(or $\eta , ~ M$) on a quasi-Fr\"obenius manifold $\em$ determines a
special geometry (or topological-antitopological fusion)
structure on it if
$$D_aC_b=D_bC_a, \eqno(1.32a)$$
$$[D_a,D_{\bar b}] = - [C_a,C_{\bar b}], \eqno(1.32b)$$
where
$$C_{\bar b} = M \overline{C_b} \bar M. \eqno(1.32c)$$
The normalized Hermitian metric
$$\tilde g_{\bar a b}d\bar z^adz^b={\gab d\bar z^adz^b \over
\gab \bar e^a e^b},$$where $e^a$ is the unity vector field, is called
generalized Zamolodchikov metric on $\em$.

{\bf Remark.} We have seen above that a compatible normalized pair $\eta$, $g$
(or $\eta$, $M$) on a manifold $\bf M$ can be encoded by a pair
$({\cal V},~G)$, where ${\cal V}$ is a real $n$-dimensional subbundle
in $T^{1,0}{\bf M}$ and $G$ is a section of the associated bundle $Q({\cal V})$
of positive definite quadratic forms on ${\cal V}$. It will be shown in
Sect.3 that the equations of special geometry have a nice geometric
reformulation in terms of the pair $({\cal V},~G)$: ${\cal V}$ is a flat
bundle (i.e. it admits a connection of zero curvature) and $G$ is a
pluriharmonic section of $Q({\cal V})$.

The equations of special geometry (together with the equations of
compatibility of $\eta$ and $M$) can be written as the following
overdetermined system of equations for the
matrix-valued function $M$:
$$\bar{\partial}_b(\partial_aM \cdot \bar M)= C_aM \bar C_b \bar M
- M \bar C_b \bar M C_a. \eqno(1.33)$$
It turns out that this system imposes a constraint for the
quasi-Fr\"obenius manifold.

{\bf Proposition 1.2.} {\it If eigenvalues of some $C_a=(c_{ab}^c(z))$
are simple and a special geometry structure on $\em$ exists then the
quasi-Fr\"obenius structure $(c_{ab}^c(z))$ is integrable.}

The proof of this proposition will be given
in the Sect.3.

Let us obtain now a "zero-curvature representation" (depending on a
spectral parameter) of the equations (1.32).

{\bf Proposition 1.3.} {\it The equations
$$\partial_a \xi = \lambda C_a \xi - \Gamma_a \xi \eqno(1.34a)$$
$$\bar{\partial}_a \xi = \lambda^{-1}  M \overline{C_a}
 \bar M \xi , \eqno(1.34b)$$
where the matrix coefficients
$$C_a=(c_{ab}^c(z)), ~ \Gamma_a=(\Gamma_{ab}^c),~ M=\Mab \eqno(1.35)$$
obey the conditions
$$M \bar M = 1, \eqno(1.36a)$$
$$M^{\rm T} \eta M = \bar{\eta}, \eqno(1.36b)$$
$$\partial_k \eta = \eta \Gamma_k + \Gamma_k^{\rm T} \eta \eqno(1.36c)$$
$$\eta C_a=C_a\trans\eta \eqno(1.36d)$$
for a holomorphic symmetric nondegenerate $\eta = (\eta_{ab})$ are
compatible identically in the spectral parameter $\lambda$ iff
the pair $\eta ,~M$ is compatible with $\Gamma_{ab}^c$ as the corresponding
connection (1.4) determining a special geometry structure on the
quasi-Fr\"obenius manifold.}

Note that the equation (1.36c) provides invariance of the compatibility
conditions
$[\partial_a ,\partial_b]=[\bar{\partial}_a, \bar{\partial}_b]=
[\partial_a, \bar{\partial}_b]=0$ with the symmetry (1.36d) and with (1.36ab).

{\bf Remark.} Compatibility of the linear equations (1.34) can be
interpreted as vanishing of the curvature of the $\lambda$-dependent
"connection"
$$\tilde D_X^{(\lambda )}Y=D_XY-\lambda X\cdot Y \eqno(1.37a)$$
$$\tilde D_{\bar X}^{(\lambda )}Y=D_{\bar X}Y-\lambda^{-1}\tau (X\cdot\tau
(Y)) \eqno(1.37b)$$
(cf.(1.27) above). For $|\lambda |=1$ the operation $\tilde D^{(\lambda )}$
determines a connection on $\bf M$ (i.e. it respects the complex
conjugation).

Proof. The compatibility $\partial_k \partial_l = \partial_l \partial_k$
implies (1.32a) and the commutativity and associativity of the algebra
$(c_{ab}^c(z))$. The compatibility $\partial_k \bar{\partial}_l =
\bar{\partial}_l \partial_k$ is equivalent to the equations
$$\bar{\partial}_l \Gamma_k = [C_k,M \bar C_l \bar M] \eqno (1.38)$$
and
$$\partial_k(M\bar C_l\bar M)= [M\bar C_l \bar M,\Gamma_k].\eqno(1.39)$$
It is enough to show that $\partial_kM+\Gamma_kM=0$.
Using $M \bar M =1$ (1.39) can be rewritten as
$$[ \bar M (\partial_k M \bar M +\Gamma_k )M, \bar C_l ]=0. \eqno(1.40)$$
Let us prove that the operator $\bar M (\partial_k M \bar M +\Gamma_k )M$
is $\bar{\eta}$-skew-symmetric. Equivalently, $\partial_kM \bar M +\Gamma_k$
should be $\eta$-skew-symmetric. Indeed,
$$0=\partial_k(M^{\rm T} \eta M )= \partial_kM^{\rm T} \eta M +
M^{\rm T} (\eta \Gamma_k + \Gamma_k^{\rm T} \eta )M +
M^{\rm T} \eta \partial_kM.$$
Multiplying by $M^{\dagger}$ and by $\bar M$ from the l.h.s. and r.h.s.
respectively one obtains the $\eta$-skew-symmetry
$$\eta (\partial_kM \bar M+\Gamma_k)+(\partial_kM\bar M+\Gamma_k)
^{\rm T} \eta=0.$$

{\bf Lemma.} {\it Let $\cal A$ be a commutative associative algebra with a
unity and $A: \cal A \to \cal A$ a linear operator such that
$$A(ab)=aA(b)=bA(a). \eqno(1.41)$$
Then $A$ is the operator of multiplication $A(a)=\alpha a, \alpha \in
\cal A$.}

Proof. Put $\alpha = A(1)$.

{\bf Corollary.} {\it If $A$ is a linear operator on a Fr\"obenius algebra
satisfying (1.41) and being  skew-symmetric w.r.t. $<~,~>$ then $A=0$.}

{}From the corollary it follows that
$$D_aM=0$$
for the connection $\Gamma_{ab}^c$. And $D_a \eta =0$ by (1.36c). The
equation (1.38) coincides with (1.32b). The Proposition is proved.

As a concequence of the Proposition we obtain

{\bf Theorem 1.} {\it Equations (1.32) of topological-antitopological
fusion are integrable}\footnote{$^{\dagger}$}{In [6, 10, 11] many integrable
reductions of (1.32) were found. Nevertheless the "Lax pair" of
[6] (coinciding with (1.34) for $\lambda = \pm 1$)
provides no possibility to solve the system in general case using an
appropriate
version of inverse spectral transform.}.

{\bf Remark.} The system (1.32) provides no restriction (additional to
(1.23), (1.36c))
for the invariant inner product $\eta$. Indeed, let us consider,
for example, the quasi-Fr\"obenius structure with the operators $C_a$ of the
form
$$c_{ab}^c=\delta_a^c \delta_{ab} \eqno(1.42)$$
(in the given coordinates). Then the invariant inner product $\eta$ should
be diagonal in these coordinates. Any diagonal holomorphic matrix $\eta '$
determines another invariant inner product. The gauge transformation (1.17)
with
$$A=\sqrt{\eta ' \eta^{-1}} \eqno(1.43)$$
transforms the special geometry structure for $c_{ab}^c, \eta$ to a
special geometry structure for  $c_{ab}^c, \eta '$.

{\bf Example.} Let us consider special geometry structures on the "trivial"
Fr\"obenius manifold $c_{ij}^k=const.,~ \eta_{ij}=const.$. In the
nonnilpotent case one can consider the direct sum of $n$ copies of
the 1-dimensional Fr\"obenius algebra
$$c_{ij}^k=\delta_i^k\delta_{ij},~ \eta_{ij}=\delta_{ij}. \eqno(1.44)$$
The first part (1.32a) of the equations reads
$$[\Gamma_i,C_j]=[\Gamma_j,C_i], \eqno(1.45a)$$
$$\Gamma_i^{\rm T}=-\Gamma_i. \eqno(1.45b)$$
These can be solved in the form
$$\Gamma_i=[q,C_i],~ q^{\rm T}=q \eqno(1.46)$$
where the off-diagonal symmetric matrix $q=(q_{ij})$ is determined
uniquely. The system (1.32b) reads
$$\partial_i M = [C_i,q]M \eqno(1.47a)$$
$$\bar{\partial}_iq=M \bar C_b \bar M - {\rm diag}(M \bar C_b \bar M)
\eqno(1.47b)$$
where `diag' means the diagonal part of the matrix $M \bar C_b \bar M$.
The matrix $q$ also satisfies the equations
$$[\partial_iq,C_j]-[\partial_jq,C_i]+[[q,C_i],[q,C_j]]=0. \eqno (1.48)$$
This follows from (1.32a). The matrix $M=(m_{\bar i j}))$ satisfies the
constraints

$$M \trans M=1,~ M=M^{\dagger}. \eqno(1.49)$$
Also it should be positive definite. In the coordinate form the system
(1.47), (1.48) reads
$$\partial_k q_{ij}=q_{ik}q_{kj}, ~~i,~j,~k~{\rm distinct} \eqno(1.50a)$$
$$\sum_k \partial_k q_{ij}=0,~~i \neq j \eqno(1.50b)$$
$$\bar{\partial}_k q_{ij}=m_{\bar k i} m_{\bar k j}, ~~i \neq j
\eqno(1.50c)$$
$$\partial_k m_{\bar i j}=m_{\bar i k} q_{kj}, ~~k \neq j \eqno(1.50d)$$
$$\sum_k \partial_k m_{\bar i j} = 0 \eqno(1.50e)$$
(one should add also the complex conjugate equations).

In the first nontrivial case $n=2$ positive definite matrices $M$ of the form
(1.49) can be represented in the form

$$M=\left( \matrix{ \cosh {\alpha \over 2} & -i \sinh {\alpha \over 2}\cr
i \sinh {\alpha \over 2} & \cosh {\alpha \over 2}} \right)
\eqno(1.51)$$
for real $\alpha$. All the functions $q_{ij}, m_{\bar i j}$ depend on the
difference
$$z=z^1 - z^2.$$
{}From (1.50d) it immediately follows that
$$q_{12} =q_{21}=- {i \over  2} \partial \alpha ,$$
where $\partial = \partial / \partial z$. From (1.50c) one obtains the
Euclidean sinh-Gordon equation for $\alpha$
$$\partial \bar{\partial} \alpha = \sinh \alpha. \eqno(1.52)$$
Recently this equation (with the opposite sign) proved to be
important in the theory of surfaces of constant mean curvature
[14, 15, 16].

In Sect. 2 it will be shown that all special geometries on a massive
Fr\"obenius manifolds locally can be described by the system (1.50).

{\bf Remark.} Equations of compatibility of rational operator
pencils of the form
$$\partial_x\xi = \lambda C\xi -\gamma\xi \eqno(1.53a)$$
$$\partial_y\xi =\lambda^{-1}\tilde C\xi \eqno(1.53b)$$
with $n\times n$ matrix coefficients $C$, $\tilde C$, $\Gamma$ were
studied for $n>2$ in the paper [17]
as a multicomponent
generalization of the Sine-Gordon eq. (for $n=2$ (1.53) gives [18]
the Sine-Gordon equation in the light-cone variables $x$, $y$). In
[19] it was shown that these equations are
gauge equivalent to the integrable [20] equations of
principal chiral field. The idea of this gauge equivalence is important
for the constructions of Sect.3. "Multidimensionalization" of the
system (1.53) by adding dependence on higher times (i.e. isospectral
deformations of (1.53)) was considered recently in [21] without
investigating of reality constraints of the type (1.36).
\medskip

{\bf 2.Topological-antitopological fusions of a massive TFT.}
\medskip

Let us consider in more details topological-antitopological fusions
of a given massive TFT. The equations (1.32) can be represented as the
compatibility conditions of the linear system
$$\partial_{\alpha}\xi = (\lambda C_{\alpha} - [C_{\alpha}, V])\xi
\eqno(2.1a)$$
$$\bar{\partial}_{\alpha}\xi  = \lambda^{-1} (M \bar C_{\alpha} \bar M)
\xi , \eqno(2.1b)$$
$\partial_{\alpha}=\partial / \partial t^{\alpha},~
\bar{\partial}_{\alpha}=\partial / \partial \bar t^{\alpha}$,
the matrix $V=(V_{\alpha}^{\beta}(t, \bar t))$ satisfies
$$\eta V=V \trans \eta. \eqno(2.2)$$

As it was proved in [5], a canonical coordinate
system $u^1(t),\dots ,u^n(t)$ locally exists on a massive
Fr\"obenius manifold such that in these coordinates
$$c_{ij}^k=\delta_i^k \delta_{ij}, \eqno(2.3a)$$
$$\eta_{ij}=h_i^2(u) \delta_{ij} \eqno(2.3b)$$
for some analytic functions $h_i(u)$. The {\it rotation
coefficients}
$$\gamma_{ij}(u) = {\partial_j h_i(u) \over h_j(u)}, ~i \neq j,
\eqno(2.4)$$
$\partial_i = \partial / \partial u^i$, are symmetric in $i,~j$.
They satisfy the following integrable system of PDE
$$\partial_k \gamma_{ij} = \gamma_{ik} \gamma_{kj},
{}~~i,~j,~k~{\rm are~ distinct},\eqno(2.5a)$$
$$\sum_k \partial_k \gamma_{ij} = 0, \eqno(2.5b)$$
$$\gamma_{ji} = \gamma_{ij}. \eqno(2.5c)$$
The zero curvature representation of the system (2.5) has the form
[22]
$$\partial_k \psi = (\lambda E_k - [E_k, \Gamma ])\psi,~
k=1, \dots , n, \eqno(2.6a)$$
$$(E_k)_{ij} = \delta_{ij} \delta_{jk},~ \Gamma = (\gamma_{ij}). \eqno(2.6b)$$
Any solution $\gamma_{ij}(u)$ of the system (2.5) determines a TFT as follows.
Let $\psi_{i\alpha}(u), ~ \alpha = 1, \dots ,n$, be a basis of solutions
of the linear system (2.6) for $\lambda = 0$. Then
$$\eta_{ii}(u) = \psi_{i1}^2(u), \eqno(2.7a)$$
$$\eta_{\alpha \beta} = \sum_i \psi_{i \alpha}(u)\psi_{i \beta}(u),
\eqno(2.7b)$$
$${\partial t_{\alpha} \over \partial u^i} =\psi_{i 1}(u)\psi_{i \alpha}(u),
{}~~t_{\alpha}=\eta_{\alpha \beta}t^{\beta}, \eqno(2.7c)$$
$$c_{\alpha \beta \gamma}(u) = \sum_i
{\psi_{i \alpha} \psi_{i \beta} \psi_{i \gamma} \over \psi_{i 1}}.
\eqno(2.7d)$$

Let us do a gauge transformation of the linear problem (2.1) of the form
$$\varphi_i(u, \bar u, \lambda) = \psi_{i \alpha}(u)
\xi^{\alpha}(t, \bar t, \lambda ), ~u = u(t) \eqno(2.8a)$$
or, equivalently,
$$\xi^{\alpha}(t, \bar t, \lambda ) = \eta ^{\alpha \beta}
\sum_i \psi_{i \beta}(u)\varphi_i(u, \bar u, \lambda). \eqno(2.8b)$$
The functions $\psi_{i \alpha}(u)$ were determined above.

{\bf Proposition 2.1.} {\it After the substitution (2.8) the system (2.1)
transforms to the following gauge equivalent system:
$$\partial_k \varphi = (\lambda E_k - [E_k, q])\varphi \eqno(2.9a)$$
$$\bar{\partial}_k \varphi = \lambda^{-1}  m E_k \bar m \varphi
\eqno(2.9b)$$
$k=1, \dots , n$, where the symmetric off-diagonal matrix $q=(q_{ij})$ has
the form
$$q_{ij}= \gamma_{ij} - v_{ij}, \eqno(2.10a)$$
where
$$V_{\alpha}^{\beta} = \sum_{i,j} \psi_{i \alpha}\psi_j^{\beta}v_{ij},
\eqno(2.10b)$$
$$\Mab = \sum_{i,j} \bar{\psi}_{i \alpha} m_{\bar i j} \psi_j^{\beta}.
\eqno(2.10c)$$
The matrix $m$ is Hermitian, positive definite and orthogonal.}

The proof is straightforward.

As a consequence we obtain

{\bf Theorem 2.} {\it The equations (1.32) of special geometry on a
massive Fr\"obenius manifold are gauge equivalent to the equations (1.50).}

Note that the system (1.50) is universal, i.e. it does not depend on
the concrete TFT model. The gauge transformation (1.50)$\to$(1.32) is
determined only by the TFT model. So for massive models the WDVV
equations for the correlators and the equations of
topological-antitopological fusion for the ground state metric can be
decoupled.

Let us consider the trivial solution of the system (1.50):
$$q = 0, ~m = 1.\eqno(2.11)$$
This gives the following special geometry structure:
$$ds^2 = \sum |\eta_{ii}|d\bar u^i du^i =
\sum_i \bar{\psi}_{i \alpha} \psi_{i \beta} d \bar t^{\alpha} dt^{\beta}.
\eqno(2.12)$$
The curvature of the corresponding connection $D$ (see (1.4)) for
the trivial solution vanishes identically.

Solutions of the system (1.50 ) close to the trivial one can be found
from the linearized system. The leading approximation for the matrix
$m$ has the form
$$m_{\bar i j} \simeq \delta_{ij} + i \alpha_{ij}(u^i-u^j,
\bar u^i-\bar u^j), \eqno(2.13a)$$
where
$$\alpha_{ji}(u, \bar u)=-\alpha_{ij}(u,\bar u) \eqno(2.13b)$$
are real solutions to the linearized sinh-Gordon
$$\partial \bar{\partial} \alpha_{ij} = \alpha_{ij}. \eqno(2.13c)$$

Let us consider massive perturbations of topological conformal
field theory (TCFT; see [3]). They are described by a
massive Fr\"obenius manifold with a one-parameter group of automorphisms.
In the flat coordinates this group acts as scaling transformations
$$t^{\alpha} \mapsto c^{1-q_{\alpha}}t^{\alpha}, \eqno(2.14a)$$
where $q_{\alpha}$ are the charges of the primary fields,
$q_1=0$,
$$\eta_{\alpha \beta} \mapsto c^{q_{\alpha}+q_{\beta}-d}\eta_{\alpha \beta}.
\eqno(2.14b)$$
where $d$ is the dimension of the model (i.e. $\eta_{\alpha \beta} = 0$ if
$q_{\alpha}+q_{\beta} \neq d)$,
$$c_{\alpha \beta \gamma} \mapsto c^{q_{\alpha}+q_{\beta}+q_{\gamma}-d}
c_{\alpha \beta \gamma}. \eqno(2.14c)$$
In the conformal point $t=0$ the primary correlators form a graded
Fr\"obenius algebra with $q_{\alpha}$ as the weights of the generators.
In the canonical coordinates $u^i$ the group acts in the standard way
$$u^i \mapsto cu^i, \eqno(2.15a)$$
$$\gamma_{ij} \mapsto c^{-1}\gamma_{ij}. \eqno(2.15b)$$
The similarity reduction (2.15) of the system (2.5) describes the
isomonodromy deformations of the linear ODE system with rational
coefficients
$$\lambda \partial_{\lambda}\psi = (\lambda U-
[U,\Gamma ])\psi ,\eqno(2.16a)$$
$$U={\rm diag}(u^1, \dots ,u^n), \eqno(2.16b)$$
$$\Gamma = (\gamma_{ij}(u)). \eqno(2.16c)$$
The integration of the similarity reduction of (2.5), (2.15) was given in
[5]. Particularly, all the scaling dimensions $q_{\alpha},~d$
are calculated as the monodromy indices of (2.16) in the point $\lambda = 0$.
For $n=3$ this similarity reduction can be reduced to a particular case of
the Painlev\'e-VI equation; for $n>3$ this is an appropriate
high order generalisation of the Painlev\'e-VI.

Let us consider now topological-antitopological fusions of a massive
perturbation of TCFT. It is natural to assume a special geometry
structure on a scaling invariant Fr\"obenius manifold to be covariant
with respect to the group (2.14), $c=\exp i \phi$. In the massive case the
similarity reduction of the system (1.50) is obtained by adding the
equations
$$\sum_{k=1}^n (u^k\partial_k-\bar u^k \bar{\partial}_k)
 q_{ij} = -q_{ij} \eqno(2.17a)$$
$$\sum_{k=1}^n (u^k\partial_k-\bar u^k \bar{\partial}_k)m_{\bar i j}
=0. \eqno(2.17b)$$
The system (1.50), (2.17) can be reduced to a system of ODE of the order
$n(n-1)$. For $n=2$ this is equivalent to the similarity reduction
of the sinh-Gordon equation (1.52) (i.e. to a particular case of the
Painlev\'e-III equation [23]. For $n>2$ the system (1.50), (2.17)
can be considered as a high-order generalisation of the Painlev\'e-III.

{\bf Lemma.} {\it The similarity reduction (2.17) of the system (1.50) is
equivalent to the compatibility conditions of the linear problem (2.9) and of
the linear differential equation in $\lambda$
$$\lambda \partial_{\lambda}\varphi = [\lambda U -
[U,q] - \lambda^{-1}m \bar U \bar m] \varphi. \eqno(2.18)$$
}

Here $U$ has the form (2.16b), $q=(q_{ij})$.

Proof. The equation (2.18) is nothing but the similarity equation
for the function $\varphi = \varphi (u, \bar u, \lambda)$ of the form
$$[\sum (u^k\partial_k-\bar u^k \bar{\partial}_k)-
\lambda \partial_{\lambda}] \varphi = 0. \eqno(2.19)$$
Compatibility of (2.19) with (2.9) is equivalent to the similarity
equations (2.17). Lemma is proved.

{\bf Corollary.} {\it The system (1.50), (2.17) gives the isomonodromy
deformations
of the linear operator (2.18).}

Isomonodromy deformations of linear operators of the form (2.18) (for $n>2$
without any constraints of the type (1.36)) were described in
[24].
Generic self-similar solutions of (1.50), (2.17) can be
found easily using an appropriate
Riemann boundary value problem (see [23]). Here we study more
thoroughly the massive case. Semi-classical arguments of [6] for
topological-antitopological fusion of Landau-Ginsburg models give rise to
specification of $n(n-1)/2$-dimensional subfamily of separatrix solutions
of (1.50), (2.17).
As it follows from the formulae of Appendix B of [6]
for such models the special geometry structure should trivialize for
$|u| \to \infty$:
$$q \to 0, ~m \to 1 ~{\rm for}~|u| \to \infty .\eqno(2.20)$$
Asymptotics of these solutions for large $u$ can be found by solving
the similarity reduction of the Helmholtz equation (2.13c). This gives
$$m_{\bar i j} \simeq \delta_{ij} + i\mu_{ij}K_0(2|u^i-u^j|)
\eqno(2.21a)$$
where
$$\mu_{ji}=-\mu_{ij} \eqno(2.21b)$$
are some real constants, $K_0(x)$ is the Bessel function (i.e.
the solution of the modified Bessel equation
$$y''+{1 \over x}y'-y=0 \eqno(2.22)$$
vanishing for $x \to \infty$). Let us construct these separatrix solutions.

To do it we formulate the following Riemann boundary value problem: to
find $n \times n$ matrix-valued functions
$\Psi_+(u, \bar u, \lambda )$ and $\Psi_-(u, \bar u, \lambda )$ analytic
in $\lambda$ in the half-planes ${\rm Re} \lambda >0$ and ${\rm Re} \lambda <0$
resp. satyisfying the following boundary conditions on the imaginary axis
(here $\rho >0$):
$$\Psi_-(u, \bar u,i \rho )=\Psi_+(u, \bar u, i\rho )S \eqno(2.23a)$$
$$\Psi_-(u, \bar u,-i \rho )=\Psi_+(u, \bar u,- i\rho )S\trans \eqno(2.23b)$$
where $S=(s_{pq})$ is a complex $n \times n$ matrix satisfying the following
conditions:
$$S \bar S=1, \eqno(2.24a)$$
$$s_{pp}=1,~~s_{pq}=0 ~{\rm for}~ \pi \leq \alpha_{pq}<2\pi \eqno(2.24b)$$
where
$$\alpha_{pq}={\rm arg}(u^p-u^q),~0\leq \alpha_{pq}<2\pi .\eqno(2.24c)$$
The functions $\Psi_{\pm}$ should obey the following normalization condition
$$\Psi_{\pm}(u, \bar u, \lambda )\exp [-\lambda U-\lambda^{-1}\bar U]
\to 1 ~{\rm for}~ \lambda \to \infty. \eqno(2.25)$$

{\bf Proposition 2.2.}{\it For sufficiently large $|u|$ there exists a unique
solution of the above Riemann b.v.p. The matrix
$$M=\lim_{\lambda \to 0} \Psi_{\pm}(u, \bar u, \lambda )
\exp [-\lambda U-\lambda^{-1}\bar U] \eqno (2.26)$$
satisfies the equations (1.33)
(for the Fr\"obenius manifold
(1.44)) being a regular function for these large
$|u|$, and $M \to 1$ for $u \to \infty$.}

Note\footnote{$^{\dagger}$}{I am acknowledged to A.Its for explaining me
how one can specify the separatrix solutions of sinh-Gordon in the
framework of the isomonodromy deformations approach.} that the solution
$M$ depends exactly on $n(n-1)/2$ parameters (i.e. on the Stokes
matrix $S$). For generic self-similar solutions of (2.9) the eigenfunction
has a jump also on a unit circle.

Proof. Let
$$\Psi_{\pm}(u, \bar u, \lambda )\exp [-\lambda U-\lambda^{-1}\bar U]=
\tilde{\Psi}_{\pm}(u, \bar u, \lambda ).$$
For the functions $\tilde{\Psi}_{\pm}(u, \bar u, \lambda )$ the Riemann
b.v.p. (2.23) reads
$$\tilde{\Psi}_-(u, \bar u,\pm i \rho )=
\tilde{\Psi}_+(u, \bar u,\pm i \rho )\tilde S^{\pm}(u, \bar u,\pm i \rho )$$
where the matrices $\tilde S^{\pm}(u, \bar u,\lambda )=
(\tilde s_{pq}^{\pm}(u, \bar u,\lambda ))$ have the form
$$\tilde s_{pq}^{+}(u, \bar u,\lambda )=\exp
[ \lambda (u^p-u^q)+\lambda^{-1}(\bar u^p-\bar u^q)]s_{pq},$$
$$\tilde s_{pq}^{-}(u, \bar u,\lambda )=\exp
[ \lambda (u^p-u^q)+\lambda^{-1}(\bar u^p-\bar u^q)]s_{qp}.$$
These matrices tend exponentially to 1 when $|u| \to \infty$
because of (2.24). The b.v.p. (2.23) is equivalent to the singular integral
equation for the function $\tilde{\Psi}_+$
$$\tilde{\Psi}_+(u, \bar u, \lambda) = 1 - \eqno(2.27)$$
$${1 \over 2\pi i}\left[
\int_0^{i\infty}{\tilde{\Psi}_+(u, \bar u, \zeta)
[1-\tilde S^+(u, \bar u, \zeta)] \over \zeta -\lambda + 0}d\zeta +
\int_{-i \infty}^{0}{\tilde{\Psi}_+(u, \bar u, \zeta)
[1-\tilde S^-(u, \bar u, \zeta)] \over \zeta -\lambda + 0}d\zeta\right] .
$$
For $|u| \to \infty$ the kernel of this equation vanishes exponentially.
This proves existence and uniqueness of the solution.

The piecewise analytic function $\Psi = (\Psi_+, \Psi_-)$ satisfies
the following identities
$$\Psi (\lambda ) \Psi \trans (-\lambda )=1 \eqno(2. 28)$$
$$\overline{\Psi (\bar{\lambda}^{-1})}=\bar M \Psi (\lambda) \eqno(2.29)$$
where $M$ is defined in (2.26). This gives
$$M\trans M=1, ~M^{\dagger} = M. \eqno(2.30)$$
The logarythmic derivatives
$$\partial_k\Psi\cdot\Psi^{-1},~ \bar{\partial}_k\Psi\cdot\Psi^{-1},~
\lambda \partial_{\lambda}\Psi\cdot\Psi^{-1}$$
are analytic in $\lambda \in {\bf C},~ \lambda \neq 0, ~ \lambda\neq\infty$.
Investigation of the analytic properties of these gives rise to infer that
they are rational functions in $\lambda$. This immediately proves that $\Psi$
is an eigenfunction of the linear problems (2.9) and (2.18), where $M$
has the form (2.26), and the matrix $V$ is determined by the formula
$$V(u,\bar u)=\lim_{\lambda \to\infty}\lambda
[1-\Psi (u,\bar u, \lambda )\exp (-\lambda U-\lambda^{-1}\bar U)].
\eqno(2.31)$$
I recall (see Prop.1.3 above) that compatibility of (2.9), (2.18)
implies that $M$ satisfies the equations (1.50), (2.17).
The proposition is proved.

Calculation of the parameters
$\mu_{ij}$ in the asymptotics (2.21) via the Stokes matrix $S$
will be given in the next publication. The
problem of global behaviour of the solutions of (1.50), (2.17)
looks more complicated (note that the Riemann b.v.p. (2.23) should
be redefined when some of the arguments of $u^p-u^q$ pass through
the real axis). To do it in the first nontrivial case $n=2$
one should use the connection formulae of [25].
\medskip

{\bf 3.Topological-antitopological fusions and pluriharmonic
maps.}
\medskip

It is wellknown that harmonic functions are the solutions $G=G(u,\bar u)$
of the equation
$$\partial \bar{\partial}G=0.$$
Pluriharmonic functions (or vector-functions) $G(z,\bar z), z=(z^1, \dots ,
z^n)$, are defined as solutions of the overdetermined system
$$\partial_k \bar{\partial}_l G=0,~~k,~l=1,\dots ,n.$$
Equivalently, the restriction of $G$ onto any holomorphic curve
$z^k=z^k(u),~k=1,\dots ,n$ should be holomorphic.

Also for any complex manifold $\bf M$ and a real Riemannian manifold $Q$
the class of pluriharmonic maps
$$G:~{\bf M} \to Q$$
is well-defined. Particularly, pluriharmonic maps of a compact complex
manifold $\bf M$ to a compact Lie groups $Q$ were studied recently
[26, 27]. They proved to have many nice features
of harmonic maps of Riemann surfaces to a compact Lie group (see
[28, 15, 29]).

In this section it will be shown that any solution of the equations (1.32)
of topological-antitopological fusion on a quasi-Fr\"obenius manifold $\bf M$
locally
determines a pluriharmonic map of $\bf M$ to the symmetric space $Q=Gl(n)/O(n)$
of $n\times n$ positive definite quadratic forms (in fact, even a loop in
the space of pluriharmonic maps ${\bf M}\to Q$). Conversely, it will be shown
that, under some additional assumptions, a pluriharmonic map ${\bf M}\to Q$
determines a family of quasi-Fr\"obenius structures together
with a special geometry structure on given ${\bf M}$.
Globally instead of pluriharmonic maps $G:~{\bf M}\to Q$ one has to
consider pluriharmonic sections $G:~{\bf M}\to Q({\cal V})$ of the bundle of
positive definite quadratic forms on a real $n$-dimensional flat subbundle
${\cal V}\subset T^{1,0}{\bf M}$ (see Sect.1 above).

Let us fix a solution of the system (1.32). Let $\Phi = \Phi_{\varphi}(z,
\bar z)$ be the fundamental matrix of the system (1.34) for $\lambda =
\exp i \varphi$ for some real $\varphi$ being normalized by the condition
$$\bar\Phi =\bar M\Phi. \eqno(3.1)$$
It is easy to see that such a normalization is compatible with (1.34).
More than that, it can be done simultaneously for any $\varphi$ so
$\Phi_{\varphi}$ is a periodic function of $\varphi$. Equivalently, the matrix
$
 M$ is factorized as
$$M=\Phi \bar \Phi^{-1}.\eqno(3.2)$$
The condition $M=M^{\dagger}$ is equivalent to reality of the symmetric
matrix
$$G=\Phi\trans \eta\Phi .\eqno(3.3)$$

Starting from this point we will consider only special geometries
with positive definite Hermitian products $g=(\gab )$. Then
$$G=\Phi^{\dagger}g\Phi \eqno(3.4)$$
is a matrix of a real positive definite quadratic form. We obtain
therefore (locally in $\bf M$) a map (depending on the parameter $\varphi$)
$$G=G_{\varphi}(z,\bar z):~{\bf M}\to Q, \eqno(3.5)$$
where $Q=Gl(n)/O(n)$ is the symmetric space of real positive definite
quadratic forms. It will be proved below that this map is
pluriharmonic.

Let us analize now the global properties of the construction.

The matrix $\Phi$ normalized as in (3.1) is determined
uniquely up to the transformations
$$\Phi \mapsto \Phi S \eqno(3.6)$$
for arbitrary real nondegenerate matrix $S$. We obtain therefore an
isomorphism between the holomorphic tangent bundle $T^{1,0}{\bf M}$ and the
complexification of some $n$-dimensional {\it real} bundle ${\cal V}\subset
T^{1,0}{\bf M}$ on
$\bf M$
$$\Phi :~{\cal V}\otimes {\bf C}\to T^{1,0}{\bf M}.\eqno(3.7)$$
Indeed, the columns of the matrix $\Phi = (\Phi_I^a) $ under holomorphic
changes of coordinates transform as holomorphic tangent vectors. In
the intersection of two coordinate charts $(z^a)$ and $(z^{a'})$ the matrices
$(\Phi_I^a)$ and $(\Phi_{I'}^{a'})$ are related by the transformation
(3.6) with a constant real $S$.
This gives the construction of the real bundle $\cal
V$. This isomorphism transforms the antiinvolution $\tau$ (with the matrix
$M$) to the identity map on $\cal V$, and the complex and Hermitian
quadratic forms $\Phi^*\eta$, $\Phi^*g$ coincide on $\cal V$ (i.e.,
they have the same real symmetric matrix $G$). We obtain that globally the
formula (3.3) determines a section of the bundle $Q({\cal V})$ of positive
definite quadratic forms on the real $n$-dimensional subbundle
${\cal V}={\rm ker}(\tau -1)\subset T^{1,0}{\bf M}$. From the construction
of the bundle $\cal V$ it immediately follows

{\bf Proposition 3.1.} {\it For
any solution of the equations (1.32) of
topological-anti\-topo\-lo\-gical
fusion the bundle ${\cal V} ={\rm ker}(\tau - 1)\subset T^{1,0}
\em$ is flat (i.e. it admits a connection of zero curvature).}

Particularly, on a simply-connected $\em$ the bundle $\cal V$
is trivial. So $G$ is a map of $\em$ to the symmetric space $Q$.
For non-simly-connected $\em$ $G$ is an automorphic map with
respect to some linear representation
$$\pi_1(\em )\to Gl(n)\eqno(3.8)$$
(twisted pluriharmonic map [31]).

Let us come back to pluriharmonic maps. The group $Gl(n)$ acts transitively
on $Q$ as follows:
$$G\mapsto S\trans G S,~S\in Gl(n). \eqno(3.9)$$
The stationary subgroup is isomorphic to $O(n)$. The invariant metric on $Q$
has the form
$${\rm tr}(G^{-1}dGG^{-1}dG).\eqno(3.10)$$
A map $G=G(z,\bar z):{\bf M}\to Q$ is called {\it pluriharmonic} if
the functional
$${i\over 4}\int {\rm tr}(G^{-1}\partial GG^{-1}\bar{\partial}G)
dud\bar u \eqno(3.11)$$
has extremum for any holomorphic curve
$$z^k=z^k(u),~k=1,\dots ,n,\eqno(3.12)$$
$$\partial = \partial /\partial u, ~\bar{\partial}=\partial /
\partial\bar u.$$
Particularly, representing
$$G=\exp \alpha \tilde G$$
where $\alpha = {1\over n}\log\det G$, $\det \tilde G=1$, one obtains
$${\rm tr}(G^{-1}\partial GG^{-1}\bar{\partial}G)=
|\partial\alpha |^2+
{\rm tr}(\tilde G^{-1}\partial \tilde G\tilde G^{-1}\bar{\partial}\tilde
G).$$
Hence $\log\det G$ is a pluriharmonic function (cf.[6]). The matrix
$\tilde G$ determines a pluriharmonic map to the irreducible symmetric space
$\tilde Q=SL(n)/SO(n)$.

The function $G(z,\bar z)$ should obey an overdetermined system of equations.
This system can be rewritten in a simple form using matrix-valued currents
$$A_k=G^{-1}\partial_kG,~A_{\bar k}=G^{-1}\bar{\partial}_kG=\bar A_k,
\eqno(3.13)$$
$$\partial_kA_l-\partial_lA_k=[A_l,A_k], \eqno(3.14a)$$
$$\bar{\partial}_l A_k={1\over 2}[A_k, A_{\bar l}]. \eqno(3.14b)$$
This system will be investigated thoroughly below. Pluriharmonicity for
sections of $Q({\cal V})$ is determined in a similar way.

The main claim of this section is

{\bf Theorem 3.} {\it Let $\Phi =\Phi(z,\bar z)$ be the fundamental matrix of
solutions of the system (1.34) for $\lambda =\exp i\varphi$ normalized by the
condition (3.1). Then $G=\Phi\trans\eta\Phi$ is a pluriharmonic section
of the bundle $Q({\cal V})$.
Conversely, let $\cal V$ be any flat real $n$-dimensional bundle on $\bf M$
satisfying (1.22bc) and $G$ a pluriharmonic section of $Q({\cal V})$. If
some of the operators $A_k =G^{-1}\partial_kG$
is semisimple (i.e. it has a simple spectrum) then the pluriharmonic map
$G:~{\bf M}\to Q$ (being defined locally)
induces on $\bf M$ a family of
integrable massive quasi-Fr\"obenius structures together with  special
geometry structures. All these quasi-Fr\"obenius structures have the same
canonical coordinates (1.30) with an arbitrary diagonal in these
coordinates holomorphic tensor $\eta$ as the invariant inner product.
Fixation of this tensor $\eta$ specifies uniquely the quasi-Fr\"obenius and
special geometry
structure on $\bf M$.}

Particularly, if the diagonal tensor $\eta$ satisfies in the canonical
coordinates to the system (2.5) (the so-called Egoroff metric) then
the above construction gives all the Fr\"obenius manifold structures
on $\bf M$ with marked atlas of canonical coordinates together with
the topological-antitopological fusions of them.

Proof. Let
$$\xi =\Phi \chi .\eqno(3.15)$$
After this gauge transform one obtains from (1.34)
$$\partial_k\chi =-{A_k\over \zeta +1}\chi \eqno(3.16a)$$
$$\bar{\partial}_k\chi = {A_{\bar k}\over \zeta -1}\chi
\eqno(3.16b)$$
where
$$A_k=2\Phi^{-1}C_k\Phi ,\eqno(3.17)$$
$$ A_{\bar k}=\bar A_k,\eqno(3.18)$$
$$\lambda ={\zeta -1\over \zeta +1}.\eqno(3.19)$$
It is easy to see that
$$A_k=G^{-1}\partial_k G \eqno(3.20)$$
for $G=\Phi\trans\eta\Phi$. Compatibility of (3.16) (identically in
the spectral  parameter $\zeta$) reads
$$[A_k, A_l]=\partial_lA_k-\partial_kA_l=0, \eqno(3.21a)$$
$$\bar{\partial}_lA_k={1\over 2}[A_k,A_{\bar l}]. \eqno(3.21b)$$
Hence the function $G(z, \bar z)$ is pluriharmonic.

To prove the converse statement one needs first to prove

{\bf Lemma.} {\it Any solution of (3.14) satisfies also (3.21).}

For the case of pluriharmonic maps to a compact Lie group
such a statement was proved in [27].

Proof. Compatibility conditions of (3.14) imply
$$[A_{\bar k},[A_p, A_q]]=0 \eqno(3.22)$$
for any $p,~q,~k$. Let us prove that $[A_p, A_q]=0$ follows from
these equations. Let us consider $G$ as a Euclidean structure
in a $n$-dimensional real vector space $V$. The operators
$A_k,~A_{\bar k}$ act in $V\otimes {\bf C}$. Let us introduce a
positive Hermitian inner product in $V\otimes {\bf C}$ by the formula
$$(v,w)=G_{IJ}\bar v^Iw^J \eqno(3.23)$$
(I recall that this is $\Phi^*g$ - see (3.4)). One has
$$A_p^{\dagger}=A_{\bar p}. \eqno(3.24)$$
Let us order anyhow pairs $\alpha =(p,q)$, $p<q$, and let
$${\cal A}_{\alpha}=[A_p,A_q],~\alpha =(p,q).\eqno(3.25)$$
It follows from (3.22), (3.24) that
$$[{\cal A}_{\alpha},{\cal A}_{\beta}^{\dagger}]=0.\eqno(3.26)$$
Particularly, the operators ${\cal A}_{\alpha}$ are normal.
Let us reduce the operator ${\cal A}_1$ to a diagonal form in
an orthonormal basis:
$${\cal A}_1={\rm diag}(\lambda_1^{(1)},\dots ,\lambda_n^{(1)}),$$
$${\cal A}^{\dagger}_1={\rm diag}
(\bar{\lambda}_1^{(1)},\dots ,\bar{\lambda}_n^{(1)}),$$
If ${\cal A}_2=(a_{IJ}^{(2)})$ then the equality $[{\cal A}_2,
{\cal A}_1^{\dagger}]=0$ reads
$$(\bar{\lambda}_I^{(1)}-\bar{\lambda}_J^{(1)})a_{IJ}^{(2)}=0.$$
Therefore
$$[{\cal A}_1,{\cal A}_2]=0.$$
Let us diagonalize these two operators simultaneously. By induction
we can reduce simultaneously to a diagonal form in an orthonormal
basis all the operators $[A_p,A_q]$.

Let $V'\subset V\otimes {\bf C}$ be a maximal subspace where all the operators
$[A_p,A_q]$ (and, therefore, $[A_{\bar p},A_{\bar q}]$) are scalars. Because
of (3.22) $V'$ is invariant with respect to all $A_k$. Then
$[A_p,A_q]\vert _{V'}=0$ since the trace of a commutator vanish. Hence all the
eigenvalues of all $[A_p,A_q]$ vanish.
Again
using their normality we obtain $[A_p,A_q]=0$. Lemma is proved.

Note that the above semisimplicity assumption has not been used in this proof.

As it has been proved pluriharmonicity of a map
${\bf M}\to Q$ is equivalent to compatibilty of the system (3.16).
Hence the equations (3.14) of pluriharmonic maps to the symmetric
space $Q$ are integrable\footnote{$^{\dagger}$}{General solution of these
equations can be obtained using the ideas of [30]}.

Now I'll try to explain the geometric idea of the final part of the proof
before to proceed to the
calculations. For simplicity let me consider the problem locally
(so all the bundles will be trivial).

Let $V$ be $n$-dimensional real vector space. The pull-back of the
Levi-Civita connection on $Q$ (for the invariant metric (3.10))
determines a complex connection $\nabla_k,~\nabla_{\bar k}$
on the trivial bundle ${\bf M}\times V\otimes {\bf C}$ (the formula
(3.37) below). The operator
$$d_G''=d\bar z^k\nabla_{\bar k} \eqno(3.27)$$
satisfies ${d_G''}^2=0$ and, therefore, it determines on ${\bf M}\times V
\otimes {\bf C}$ a structure of holomorphic vector bundle
${\cal V}\otimes {\bf C}$. The commuting operators $A_k$ act on
${\cal V}\otimes {\bf C}$. They appear to be holomorphic sections
of the bundle
$$E=T_*^{1,0}{\bf M}\otimes End({\cal V}\otimes {\bf C}).\eqno(3.28)$$
The commutativity (3.14a) can be rewritten in the form
$$A\wedge A=0,\eqno(3.29)$$
where
$$A=A_k dz^k \eqno(3.30)$$
is the section of $E$. Also the matrix-valued 1-form $A$ is closed.
Such a pair $(E,A)$ was called Higgs bundle
in [31].
It is very important that in our case there is a
Euclidean scalar product on ${\bf M}\times V$ (being specified
by the matrix $G$). It proves to be holomorphic with respect to
$d_G''$. So it determines a holomorphic nondegenerate quadratic form
$<~,~>$ on ${\cal V}\otimes {\bf C}$ being invariant for the operators
$A_k$. We obtain therefore a holomorphic family of commuting operators
being symmetric with respect to a holomorphic inner product $<~,~>$.
This looks so similar to deformation of Fr\"obenius algebras! To complete
the construction one needs to identify ${\cal V}\otimes {\bf C}$ and
$T^{1,0}{\bf M}$ (i.e. to construct an isomorphism $\Phi$ - see
(3.7)). Here the semisimplicity assumption is essential. We construct
localy a basis of holomorphic sections $v_1,\dots ,v_n$ of
${\cal V}\otimes {\bf C}$ and such a coordinate system $u^1,\dots ,
u^n$ on $\bf M$ that
$$Av_k=du^kv_k \eqno(3.31)$$
(no summation over the repeated indices here!). The isomorphism we need
is constructed then as follows
$$v_k\mapsto \partial /\partial u^k.\eqno(3.32)$$

Let us proceed now to the detail proof.

Let $\omega = \omega_kdz^k$ be a 1-form such that
$$A_kv=\omega_kv \eqno(3.33)$$
for some common vector function $v=v^I(z,\bar z).$
The genericity assumption provides existence of $n$ linearly independent
forms $\omega^1,\dots ,\omega^n$ being the "weights" (3.33) of the
commutative algebra $A_1,\dots , A_n$. Let $v_1,\dots , v_n$ be the
corresponding eigenvectors. Let
$$<v,w>=G_{IJ}v^Iw^J \eqno(3.34)$$
be a symmetric nondegenerate inner product in $V$ (and in $V\otimes {\bf C}$).
The operators $A_1,\dots ,$ $A_n$ are symmetric with respect to this inner
product. The standard consequence of this fact is that
$$<v_I,v_J>=0~{\rm for}~I\neq J,\eqno(3.35)$$
$$<v_I,v_I>\neq 0. \eqno(3.36)$$
Let us prove that all the weights $\omega$ are holomorphic forms on $\bf M$.
Indeed,
$$0=\bar{\partial}_l(A_kv-\omega_kv)=
{1\over 2}[A_k, A_{\bar l}]v+(A_k-\omega_k)\bar{\partial}_lv-
\bar{\partial}_l\omega_kv.$$
Multiplying by $v$ one obtains
$$\bar{\partial}_l\omega_k<v,v>=
{1\over 2}<v,[A_k,A_{\bar l}]v>=0.$$
The next step: to prove that the weights are closed forms. The proof:
$$\partial_l(A_kv)-\partial_k(A_lv)=(\partial_l\omega_k-
\partial_k\omega_l)v+
\omega_k\partial_lv-\omega_l\partial_kv.$$
After scalar multiplication by $v$ one obtains
$$(\partial_l\omega_k-\partial_k\omega_l)<v,v>=0$$

It is convenient to introduce the conection in the trivial bundle
${\bf M}\times V$
$$\nabla_kv=\partial_kv+{1\over 2}A_kv,~
\nabla_{\bar k}v=\bar{\partial}_kv+{1\over  2}A_{\bar k}v \eqno(3.37)$$
(the pull-back of the Levi-Civita connection on $Q$).
It is compatible with the scalar product (3.34):
$$\partial_k<v,w>=<\nabla_kv,w>+<v,\nabla_kw>,$$
$$\bar{\partial}_k<v,w>=<\nabla_{\bar k}v,w>+<v,\nabla_{\bar k}w>$$
for any vectors $v,~w$.

Let us prove that the eigenvectors can be normalized in such a way
that
$$\nabla_{\bar k}v=0 \eqno(3.38)$$
(i.e. that they are holomorphic sections of ${\cal V}\otimes {\bf C}$
w.r.t. (3.37)). Indeed, if $v$, $v'$ are two eigenvectors with the
eigenvalues $\omega$, $\omega '$, then from
$$0=\bar\partial_l(A_kv-\omega_kv)=
{1\over 2}[A_k,\bar A_l]v+(A_k-\omega_k)\bar{\partial}_lv$$
after multiplication by $v'$ one obtains
$$(\omega_k'-\omega_k)<v',\bar\partial_lv+{1\over 2}\bar A_lv>=0$$
Hence
$$\nabla_{\bar l}v=f_{\bar l}v$$
for some functions $f_{\bar l}$. It is easy to see that the 1-form
$f_{\bar l}d\bar z^l$ is closed. After renormalisation of $v$ we
obtain (3.38).

The norms $<v,v>$ of the normalized eigenvectors are holomorphic
functions:
$$\bar\partial_l<v,v>=2<\nabla_{\bar l}v,v>=0.$$
Let $v_1,\dots ,v_n$ be any normalized basis of the eigenvectors   (3.33).
Let
$$\eta_{ij}=<v_i,v_j> \eqno(3.39)$$
(a diagonal holomorphic matrix). Then we put
$$\Phi =\hat v^{-1} \eqno(3.40)$$
where the coordinates of the eigenvectors $v_1,\dots ,v_n$ are written
as the columns of the matrix $\hat v$. One has
$$G=\Phi\trans\eta\Phi ,\eqno(3.41a)$$
$$A_k=2\Phi^{-1}C_k\Phi \eqno(3.41b)$$
where
$$2C_k={\rm diag}(\omega^1_k,\dots ,\omega^n_k).\eqno(3.42)$$
To define the connection $\Gamma$ let us represent the covariant derivatives
$\nabla_kv_i$ as linear combinations of the basis vectors
$$\nabla_kv_i=\Gamma_{ki}^jv_j.\eqno(3.43)$$
The connection $ \Gamma_{ki}^j$ is compatible with $\eta_{ij}$ as it follows
from the definition. It is easy to see that after the gauge transformation
$$\chi =\hat v\xi$$
and the substitution (3.41) - (3.43) the linear problem (3.16) transforms to
(1.34), where
$$M=\Phi\bar\Phi^{-1}.$$
We are to prove only that the operators $C_k$ determine a closed algebra
with a unity. Indeed, since the forms $\omega^i$ are closed, they
have the form
$$\omega^i=2du^i\eqno(3.44)$$
for some functions $u^i$. In the coordinates $u^1,\dots ,u^n$ the
operators $C_i$ become the matrix unities
$$(C_i)_p^q=\delta_i^q\delta_{ip}.\eqno(3.45)$$
This completes the proof of the theorem.

Note that Proposition 1.2 follows from (3.45).
\medskip
{\bf Appendix. WDVV equations for massive TFT models and correlators
of the impenetrable Bose gas.}
\medskip

After the main body of this paper has been written I was informed by S.Cecotti
that he also obtained (in a joint work with C.Vafa) the universal form (1.50),
(2.17) of the equations of topological-antitopological fusion in TCFT
for the particular case of topological minimal models. In the derivation
[32] Cecotti and Vafa also used the canonical coordinates (2.3) coinciding
[5] for the case of minimal models with critical values of Landau - Ginsburg
superpotential. Moreover, they observed [32] that these equations
coincide\footnote{$^{\dagger}$}{I am
acknowledged to S.Cecotti for explanation of intrinsic
physical reasons of this coincidence.} with the equations [33] for multipoint
correlators in 2D Ising model.

In this appendix I will show that the equations [34] for multipoint
correlators in the impenetrable Bose gas (as functions of distances)
are in close relations with WDVV equations (written in the canonical
coordinates
(2.3)) for a massive TFT model
with even number of primaries.

I recall that the Hamiltonian of the one-dimensional non-relativistic Bose
gas [35] has the form
$$H=\int_{-\infty}^{+\infty}(\partial_z\psi^+\partial_z\psi
+\psi^+\psi^+\psi\psi - h\psi^+\psi)dz \eqno(A.1)$$
where $\psi (z)$, $\psi^+(z)$ are canonical Bose fields,
$[\psi (z),\psi^+(z')]=\delta (z-z')$, $h$ is a chemical potential.
For the case of impenetrable bosons the coupling constant being
$c=+\infty$. The mean value of an operator ${\cal O}$ at temperature $T>0$
is defined in a standard way
$$<{\cal O}>_T ={\rm tr}({\cal O}\exp (-H/T))/{\rm tr}(\exp (-H/T)).
\eqno(A.2)$$
Multi-point correlators are defined as the mean values of product of the
operators $\psi^+(z^+_i)$, $\psi(z^-_j)$ for different real
$z_i$, $z_j$:
$$<\psi^+(z_1^+)\dots \psi^+(z_N^+)\psi(z_1^-)\dots \psi(z_N^-)>_T=
T^{N/2}G_N(x^+,x^-,t) \eqno(A.3)$$
for some function $G_N$ of the variables
$$x^{\pm}=(x^\pm_1,\dots ,x^\pm_N),~x^\pm_i=z_i\sqrt T, ~
i=1,\dots ,N,$$
$$t=h/T. \eqno(A.4)$$
This function can be represented in the form [34]
$$G_N(x^+,x^-,t)=(1/4)^N(-1)^{[(N+1)/2]}$$
$$\times \prod_{j<k}{\rm sign} (x_k^+-x_j^+){\rm sign}
(x_k^--x_j^-)$$
$$\times \det_N(V_{lm}(x,t,\kappa ))\Delta (x,t,\kappa )
)|_{\kappa = {2\over \pi}}. \eqno(A.5)$$
Here $\det_N(V_{lm})$ means determinant of the $N\times N$ minor
of the form $1\leq l \leq N$, $N+1\leq m\leq 2N$.
The real functions $V_{lm}(x,t,\kappa )$ and $\Delta (x,t,\kappa )$
are determined in terms of the linear integral operator
$$({\rm K}f)(\lambda )=\int_{-\infty}^\infty \sum_{m=1}^{2N}
(-1)^m{e_m^+(\lambda )e_m^-(\mu )\over 2i(\lambda - \mu)}
f(\mu )d\mu \eqno(A.6)$$
where
$$e_m^\pm (\lambda )=\sqrt {\theta (\lambda )}\exp (\pm i\lambda x_m),
\eqno(A.7)$$
$\theta (\lambda )$ is the Fermi weight
$$\theta (\lambda )=[1+\exp (\lambda^2-t)]^{-1}\eqno(A.8)$$
(dependence of $\theta (\lambda )$ on $t$ and of $e_m^\pm (\lambda )$
on $x$, $t$ is suppressed in the formulae). Namely,
$$V_{lm}(x,t,\kappa )=(-1)^l\kappa \int_{-\infty}^\infty
e_l^-(\mu )f_m^+(\mu )d\mu ,\eqno(A.9)$$
where the functions $f_m^\pm (\mu )$ (depending also on $x$, $t$,
$\kappa$) are determined by the formulae
$$f_m^+-\kappa {\rm K}f_m^+=e_m^+ \eqno(A.10a)$$
$$f_m^--\kappa {\rm K}^\dagger f_m^-=e_m^-. \eqno(A.10b)$$
The function $\Delta (x,t,\kappa )$
$$\Delta (x,t,\kappa )=\det (1-\kappa {\rm K}).\eqno(A.11)$$
The functions $V_{lm}$ obey the following integrable system of PDE [34]
$$\partial_kV_{lm}={1\over 2}V_{lk}V_{km},~k\neq m,l\eqno(A.12a)$$
$$\sum_{k=1}^{2N}\partial_kV_{lm}=0,\eqno(A.12b)$$
$$V_{ml}=(-1)^{l+m}V_{lm},\eqno(A.12c)$$
$$\sum V_{mm}=0\eqno(A.12d)$$
as functions of the $2N$-component vector $x=(x^+,x^-)=(x_1,\dots ,
x_{2N})$, $\partial_k=\partial /\partial x_k$. Equations containing the
"time" derivative are [ibid]
$$\partial_t(\partial_l-\partial_m)V_{lm}+(x_l-x_m)V_{lm}
+(\partial_tV_{mm}-\partial_tV_{ll})V_{lm}$$
$$+{1\over 2}\sum_{p\neq l,m}(V_{lp}\partial_tV_{pm}
-V_{pm}\partial_tV_{lp})=0.\eqno(A.13)$$
Dependence of the determinant (A.11) on $x$, $t$ is specified by the
equations
$$\partial_m\log\Delta =-{1\over 2}V_{mm}\eqno(A.14a)$$
$$\partial_t\log\Delta =-{1\over 2}\sum_{m=1}^{2N}
x_m\partial_tV_{mm}+{1\over 4}\sum (\partial_tV_{lm})(
\partial_tV_{ml}).\eqno(A.14b)$$
After substitution
$$V_{lm}=(-1)^l\gamma_{lm},\eqno(A.15a)$$
$$x_l=2(-1)^lu^l\eqno(A.15b)$$
one immediately obtains that the $t$-independent part (A.12) of the
system for the off-diagonal part of the matrix $V$ coincides with the
system (2.5) for $n=2N$ (i.e. with WDVV eqs. in the canonical
coordinates $u^l$). The functions $V_{mm}$ and $\log\Delta$ are determined
by the off-diagonal terms $V_{lm}$ from the equations (A.12) and (A.14).

We obtain that multipoint correlators of the impenetrable Bose-gas
determine a two-parameter family of massive TFT models (depending on
$t$ and $\kappa$) with even number of
primaries. The zero-temperature limit of these correlators can be expressed via
Painlev\'e transcendents of the fifth kind (for $N=1$) and their high-order
generalisations [36]. It is interesting that this limit does not coincide
with TCFT, where all the correlators are given in terms of the Painlev\'e-VI
transcendents and their high-order generalisations.

\vskip 2cm
{\bf Acknowledgments.}
\medskip

I am acknowledged to S.Cecotti for explanation of some important points
of the paper [6]. I wish to thank A.Its for help in using
the isomonodromy deformations method. I am grateful to E.Witten for paying
my attention to the papers [31]; this was useful for giving
the Sect.3 its final shape.

\vfill\eject
{\bf References.}
\medskip
\item{1.} E.Witten, Comm. Math. Phys. {\bf 118}
(1988) 411; Nucl. Phys. {\bf B340}
(1990) 281.
\medskip
\item{2.} R.Dijkgraaf and E.Witten, Nucl. Phys. {\bf B342} (1990) 486.
\medskip
\item{3.} R.Dijkgraaf, E.Verlinde, and H.Verlinde, Nucl. Phys.
{\bf B352} (1991) 59.
\medskip
\item{4.} E.Witten, Surv. Diff. Geom. {\bf 1} (1991) 243;
\item{}R.Dijkgraaf, {\it Intersection
Theory, Integrable Hierarchies and Topological Field Theory}, Preprint
IASSNS-HEP-91/91, December 1991.
\medskip
\item{5.} B.Dubrovin, {\it Integrable Sys\-tems in Topo\-logi\-cal Field
The\-ory},
Preprint INFN/AE-92/01. To be published in Nucl. Phys. {\bf B}.
\medskip
\item{6.} S.Cecotti, C.Vafa, Nucl. Phys. {\bf B367} (1991) 359.
\medskip
\item{7.} A.Strominger, {\it Topology of
Superstring Compactification,} in "Unified String Theory" eds. Green and
Gross (World Scientific Publishing Co. 1986);
\item{}N.Seiberg, Nucl. Phys. {\bf B303} (1988)
286;\item{}P.Candelas, P.S.Green and T.Hubsch, {\it Connected Calabi-Yau
compactifications}, Proceedings of Maryland Superstring '88 Workshop;
\item{}S.Cecotti, S.Ferrara, and L.Girardello, Int. J. Mod. Phys.
{\bf A4} (1989) 2475;
\item{}S.Cecotti, Commun. Math. Phys. {\bf 131} (1990) 517;
\item{}S.Cecotti, Comm. Math. Phys.
{\bf 124} (1989) 23;
\item{}S.Ferrara and A.Strominger, {\it $N=2$ Spacetime Supersymmetry
and Calabi-Yau Moduli Space}, presented at Texas A$\&$M University,
Strings '89 Workshop;
\item{}A.Strominger, Comm. Math. Phys. {\bf 133} (1990) 163;
\item{}P.Candelas and X.C.de la Ossa, {\it Moduli Space of Calabi-Yau
Manifolds}, University of Texas Report UTTG-07-90;
\item{}L.Castellani,
R.D'Auria and S.Ferrara, Class. Quantum Grav. {\bf 7} (1990) 1767.
\medskip
\item{8.} S.Cecotti, Int. J. Mod. Phys. {\bf A6} (1991) 1749;
\item{}S.Cecotti,
Nucl. Phys. {\bf B355} (1991) 755.
\medskip
\item{9.} A.B.Zamolodchikov, JETP Lett. {\bf 43} (1986) 731.
\medskip
\item{10.} S.Cecotti and C.Vafa, {\it Exact Results for Supersymmetric
Sigma Models}, Preprint HUTP-91/A062.
\medskip
\item{11.} S.Cecotti and C.Vafa, {\it Mas\-sive Orbi\-folds},
Pre\-print HUTP-92/A013 and SISSA 44/\-92/\-EP.
\medskip
\item{12.} W.A.Leaf-Hermann,
{\it From Here to Criticality: Renormalization Group Flow
Between Two Conformal Field Theories}, Preprint HUTP-91/A061.
\medskip
\item{13.} S.Kobayashi and K.Nomizu, {\it Foundations
of Differential Geometry}, Vol.2, Intersci. Publ., New York/London, 1963.
\medskip
\item{14.} U.Pinkall and I.Sterling, Ann. Math. {\bf 130}
(1989) 407.
\medskip
\item{15.} N.S.Hitchin, J. Diff. Geom. {\bf 31} (1990) 627.
\medskip
\item{16.} A.I.Bobenko, Math. Ann. {\bf 290} (1991) 209.
\medskip
\item{17.} A.S.Budagov and L.A.Takhtadjan, Dokl. Akad. Nauk SSSR
{\bf 235} (1977) 805,
English translation Soviet Phys. Dokl. {\bf 22} (1977) 428.
\medskip
\item{18.} M.Ablowitz et al., Phys. Rev. Lett. {\bf 30} (1973) 1262;
\item{}V.E.Zakharov,
L.Takhtadjan and L.D.Faddeev, Dokl. Akad. Nauk SSSR {\bf 219} (1974) 1334,
English transl. in Soviet Phys. Dokl. {\bf 19} (1974/75).
\medskip
\item{19.} V.E.Zakharov and A.V.Mikhailov, J.Eksper. Teoret.
Fiz. {\bf 74} (1978) 1953, English transl. in Soviet Phys. JETP
{\bf 47} (1978).
\medskip
\item{20.} K.Pohlmayer, Commun. Math. Phys. {\bf 46} (1976) 207.
\medskip
\item{21.} M.Ablowitz, R.Beals, and K.Tenenblat, Stud. Appl. Math. {\bf 74}
(1986) 177.
\medskip
\item{22.} B.Dubrovin, Funct. Anal. Appl. {\bf 11} (1977) 265.
\medskip
\item{23.} A.R.Its, V.Yu.Novokshenov, {\it The
isomonodromic Deformation Method in the Theory of
Painlev\'e equations}, Lecture Notes in Mathematics {\bf 1191},
Springer Verlag, Berlin 1986.
\medskip
\item{24.} K.Ueno, Proc. Jpn. Acad. Sci. {\bf A56} (1980) 97.
\medskip
\item{25.} B.M.McCoy, C.A.Tracy, T.T.Wu, J. Math. Phys. {\bf 18} (1977) 1058;
\item{}V.Yu.Novokshenov, Funct. Anal. Appl. {\bf 20} (1986) 113.
\medskip
\item{26.} Y.Ohnita, J. London. Math. Soc. {\bf (2) 35}
(1987) 563;
\item{}S.Udagawa, Proc. London Math. Soc. {\bf (2) 37} (1988) 375.
\medskip
\item{27.} Y.Ohnita and G.Valli, Proc. London Math. Soc. {\bf (3) 61}
(1990) 546.
\medskip
\item{28.} K.Uhlenbeck, J.Diff. Geom. {\bf 30}
(1989) 1.
\medskip
\item{29.} J.Eells and L.Lemaire,
Bull. London Math. Soc. {\bf 10} (1978) 11; ibid., {\bf 20} (1988) 385.
\medskip
\item{30.} I.M.Krichever,
Dokl. Acad. Nauk SSSR {\bf 253} (1980) No.2, English transl. Soviet Math.
Dokl. {\bf 22} (1980) 79.
\medskip
\item{31.} N.J.Hitchin, Proc. London Math. Soc. {\bf (3)
55} (1987) 59;
\item{}S.K.Donaldson, ibid., p.127;
\item{}C.T.Simpson, J. Amer. Math. Soc. {\bf 1} (1988) 867.
\medskip
\item{32.} S.Cecotti, Lectures at Spring School and Workshop in String
Theory, Trieste, March 30 - April 10, 1992.
\medskip
\item{33.} M.Sato, T.Miwa, and M.Jimbo, Publ. RIMS {\bf 14} (1978) 223;
{\bf 15} (1979) 201, 577, 871; {\bf 16} (1980) 531.
\medskip
\item{34.} A.R.Its, A.G.Izergin, and V.E.Korepin, Comm. Math. Phys.
{\bf 129} (1990) 205.
\medskip
\item{35.} Mathematical Physics in One Dimension. E.H.Lieb, D.S.Mattis
(eds.). New York: Academic Press 1966.
\medskip
\item{36.} A.Jimbo, T.Miwa, Y.Mori, M.Sato, Physica {\bf 1D} (1980) 80.
\medskip
\hfill March - April 1992

\vfill\eject\end